\documentclass[11pt,a4paper,twoside]{article}

\usepackage{epsfig}
\usepackage[latin1]{inputenc}
\def\be{\begin{equation}}
\def\ee{\end{equation}}
\def\ba{\begin{array}}
\def\bacc{\begin{array} {cc}}
\def\ea{\end{array}}
\def\bea{\begin{eqnarray}}
\def\eea{\end{eqnarray}}
\def\bd{\begin{displaymath}}
\def\ed{\end{displaymath}}

\begin{document}

\begin{center}

{\Large\bf Relaxing Lorentz invariance in general perturbative anomalies}

\vspace{1cm}
{\large A. Salvio}\\

\vspace{.6cm}

{\it {Institut de Th\'eorie des Ph\'enom\`enes Physiques,
\\EPFL, CH-1015 Lausanne, Switzerland\\Email:
alberto.salvio@epfl.ch}}

\end{center}

\vspace{1cm}

\begin{abstract}

We analyze the role of Lorentz symmetry in the perturbative non-gravitational anomalies for a single family of fermions. The theory is assumed to be translational invariant, power-counting renormalizable and based on a local action, but is allowed to have general Lorentz violating operators. We study the conservation of global and gauge currents associate with general internal symmetry groups and find, by using a perturbative approach, that Lorentz symmetry does not participate in the clash of symmetries that leads to the anomalies. We first analyze the triangle graphs and prove that there are regulators for which the anomalous part of the Ward identities exactly reproduces the Lorentz invariant case. Then we show, by means of a regulator independent argument, that the anomaly cancellation conditions derived in Lorentz invariant theories remain necessary ingredients for anomaly freedom.

\end{abstract}

\newpage

\section{Introduction} \label{Intro}

The understanding of anomalies in quantum field theories has important applications both because it provides a strict criterion
to select consistent gauge theories and because it represents a way to explain a number of experimental facts (for the original works see \cite{Bell:1969ts}).

The explicit calculation of the anomalous Ward identities in standard theories shows that the anomaly manifests itself as the impossibility to define two or more conserved currents in the quantum theory rather than the breaking of a single symmetry. The vector and axial symmetries cannot be simultaneously implemented  taking into account quantum corrections, but it is possible to define the theory in a way that, for example, the vector symmetry is preserved. From this point of view an interesting conceptual question is weather or not the space-time symmetry that we observe in nature participates in this clash of symmetries; in other words, is the anomaly the impossibility to define a quantum relativistic theory with conserved vector and axial currents or does  Lorentz symmetry behave as a spectator in this clash? The aim of the paper is to address this problem in a broad class of Lorentz violating theories, focusing on those that can be treated perturbatively and do not include gravitational interactions.

On the other hand, the investigation of quantum field theories that do not possess the relativistic symmetry has attracted much interest among particle physicists during the last decades. One of the most fascinating motivations is the possibility to interpret these models as effective descriptions of more fundamental theories where gravity is consistently included and the breaking of Lorentz invariance occurs spontaneously \cite{Wald:1980nm}. Relaxing Lorentz invariance might also open further ways to address phenomenological problems \cite{Bertolami:1996cq} and for these reasons much work has been done to investigate conceptual issues of quantum field theory with Lorentz violation, for example causality, stability \cite{Kostelecky:2000mm} and renormalization \cite{Kostelecky:2001jc,Colladay:2006rk}. Among the most popular frameworks, which extend the Standard Model (SM) by incorporating Lorentz violating operators, there are the works by Coleman and Glashow \cite{Coleman:1998ti} and Colladay and Kostelecky \cite{Colladay:1996iz,Colladay:1998fq}, where quite general assumptions like locality, translational invariance and power-counting renormalizability are made.

In this paper, by making those assumptions, we provide a derivation of the perturbative anomalies allowing general (but small) Lorentz violating operators. For the sake of definiteness, here we focus on the case of a single family of fermions or at least on the case in which the Lorentz violating parameters are diagonal in the family space.  Our strategy is to deal with fermions coupled to gauge fields, which can be either dynamical or non-dynamical, therefore describing both the anomalies of gauge and global currents. The (one-loop) 3-point (Green) functions involving three currents are analyzed and the anomalous parts are proved to be independent of all the Lorentz violating parameters, which are compatible with the gauge symmetries at the classical level, by using a certain momentum cut-off regularization. Although we have more freedom to change the form of the anomalies in this framework (as Lorentz non-invariant counterterms are allowed), we  show that this is not sufficient to eliminate the anomaly by adding local terms to the Lagrangian, which are polynomial in the gauge fields and their derivatives. Therefore, the anomaly cancellation conditions turn out to be very stable.

This paper is organized as follows. In Section \ref{model} we describe in detail our class of  Lorentz violating fermion models coupled to general gauge fields. We then consider in Section \ref{fermionredef} the 3-point functions for the currents and show, in particular regularizations, that their anomalous part reproduces the analogous quantities in the Lorentz invariant case. In Section \ref{regulator-indep}, we provide a regulator independent argument, which shows that the anomaly cancellation conditions remain at least as strong as in the Lorentz invariant limit. We end with some conclusions in Section \ref{Concl}.

\section{The fermion sector} \label{model}

We consider a family of fermions $\psi$ in a general representation\footnote{We take as part of the definition of family the property that the representation can be made by one or more than one irreducible representations (irreps), but the irreps are all different.} of a (compact Lie) group $G$. In the following we will write an action that contains a certain number of Lorentz violating terms. Let us assume the corresponding parameters to be small (in a sense that will be clarified later on) and the corresponding theory to be perturbative with respect to (w.r.t.) them. This assumption allows us to consider $\psi$ as an object with four {\it spinorial} components, as a discrete quantity cannot be changed by a small perturbation.  The infinitesimal action of $G$ on $\psi$ is
\be \delta \psi = i \Omega \psi\equiv i \Omega^b \left(t^b_L P_L  + t^b_R P_R\right)\psi, \label{psitransf}\ee
where $\Omega^b$ represent the group transformation parameters, $P_{L(R)}\equiv \left(1\pm \gamma_5\right)/2$ is the projector on the left-handed (right-handed) subspaces, $\gamma_5$ is defined as in Lorentz invariant theories and $t^b_{L(R)}$ are the hermitian generators in the left-handed (right-handed) representation.

We imagine that each generator of $G$ corresponds to a gauge field $A_{\mu}^b$, $\mu=0,1,2,3$, but we do not require each gauge field to be dynamical: in this way we can study both the anomalies associate with gauge currents and those associate with global currents. The action of $G$ on $A_{\mu}^b$ and the covariant derivative of the fermion are respectively defined by 
\be \delta A_{\mu}^b =  f^{c d b} \Omega^d A_{\mu}^c + \partial_{\mu} \Omega^b, \label{Atransf}\ee
where $f^{c d b}$ are the structure constants of $G$, satisfying $[t^b_L,t^c_L]=if^{bcd}t^d_L$, and 
\be D_{\mu}\psi \equiv \left[\partial_{\mu} -iA_{\mu}^b\left(t^b_L P_L  + t^b_R P_R\right) \right]\psi\ee
like in Lorentz invariant theories, as their form is only consequence of gauge invariance and therefore is insensitive to any Lorentz violation. Here we only consider chiral representations: 
\be t^b_L \neq t^b_R, \quad \mbox{for some $b$}.\label{chirality}\ee
This is indeed the only case when anomalies can appear in Lorentz invariant theories.

All the ingredients introduced so far are also present in standard theories.
We now want to write a (classical) action for $\psi$ in the $A^b_{\mu}$ background, which involves Lorentz violating terms. By following the works of Coleman and Glashow \cite{Coleman:1998ti} and Colladay and Kostelecky \cite{Colladay:1996iz,Colladay:1998fq}, we assume the following properties:
\begin{itemize}
\item Locality,
 \item Translational invariance,
\item Power-counting renormalizability (operators with dimension greater than four are not allowed).
\end{itemize}
These requirements tell us that the general form\footnote{We adopt the signature $\eta_{\mu \nu}=$ diag$(+1,-1,-1,-1)$.} of the action is \cite{Colladay:1998fq}
\be S= \int d^4x \left(i\,\overline{\psi}\Gamma^{\mu} D_{\mu} \psi - \overline{\psi} M\psi \right),\label{action}\ee
where $\Gamma^{\mu}$, $\mu=0,1,2,3$ and $M$ are general\footnote{An additional term of the form $m'\gamma_5$, where $m'$ is a constant, can be added to (\ref{Mdef}), but this may be removed from the action via a chiral transformation.}  constant $4\times 4$ matrices:
\bea \Gamma^{\mu}&\equiv &c^{\mu}_{\,\,\,\nu}\gamma^{\nu} +d^{\mu}_{\,\,\,\nu}\gamma_5 \gamma^{\nu} + e^{\mu} + i f^{\mu} \gamma_5 +\frac{1}{2} g^{\mu \nu \rho} \sigma_{\nu \rho}, \\
M&\equiv &m + \frac{1}{2}H^{\mu \nu} \sigma_{\mu \nu} + a_{\mu}\gamma^{\mu} +b_{\mu} \gamma_5 \gamma^{\mu}. \label{Mdef}\eea
Here $\gamma^{\mu}$ are the usual Dirac matrices\footnote{Also $\gamma_5= i\gamma_0 \gamma_1 \gamma_2 \gamma_3$ and $\sigma_{\mu \nu}=i[\gamma_{\mu},\gamma_{\nu}]/4$.} and we have introduced the Lorentz violating parameters 
\be c^{\mu}_{\,\,\,\nu}-\delta^{\mu}_{\nu}, d^{\mu}_{\,\,\,\nu}, e^{\mu}, f^{\mu}, g^{\mu \nu \rho}, H^{\mu \nu}, a_{\mu}, b_{\mu}.\label{LVparam}\ee
If the fermion representation is made by more than one irreps, the parameters in (\ref{LVparam}) are different for different irreps; we understand here an additional index labeling different irreps.
The experimental limits require that, in a frame in which the earth is not relativistic, all the quantities in (\ref{LVparam}) are very small, in the sense that  $c^{\mu}_{\,\,\,\nu}-\delta^{\mu}_{\nu}, d^{\mu}_{\,\,\,\nu}, e^{\mu}, f^{\mu}, g^{\mu \nu \rho}<<1$ and  $H^{\mu \nu}, a_{\mu}, b_{\mu}<< m$ \cite{Kostelecky:2000mm}. In this paper we always work in such a frame. This supports the validity of the perturbation theory w.r.t. these parameters, which we use throughout the paper. Also the parameters in (\ref{LVparam}) and $m$ are real\footnote{In our conventions $\overline{\psi}\equiv \psi^{\dagger}\gamma^0$ and $\left(\gamma^{\mu}\right)^{\dagger}=\gamma^0 \gamma^{\mu} \gamma^0$.} as a consequence of $S^{\dagger}=S$. 

The consistency of this model (in the free field case, $A_{\mu}^b=0$) has been investigated in \cite{Kostelecky:2000mm}. There it is shown that inconsistencies emerge at very high energies or equivalently in frames that move at very high speed w.r.t. earth-based laboratories. These energies (or equivalently boosts) are at a very high scale $\Lambda$ where the spontaneous symmetry breaking of Lorentz invariance occurs. For example it is conceivable, but not obligatory, that $\Lambda$ is the Planck scale. Therefore, the model at hand should be considered as a low energy effective description. From an effective field theory point of view we expect \cite{Kostelecky:2000mm} $c^{\mu}_{\,\,\,\nu}-\delta^{\mu}_{\nu}, d^{\mu}_{\,\,\,\nu}, e^{\mu}, f^{\mu}, g^{\mu \nu \rho}$ to be at most of order $m/\Lambda$ and $H^{\mu \nu}, a_{\mu}, b_{\mu}$ to be at most of order $m^2/\Lambda$ and therefore these parameters are tiny if $m$ is identified with the mass of the observed fermions.

In the following we will not assume $-\overline{\psi}M\psi$ to be invariant under (\ref{psitransf}); in this way our analysis will be applicable also to those theories, like the minimal SM, where the fermion masses emerge from the spontaneous symmetry breaking of a gauge symmetry. However, we do assume the first term in (\ref{action}) to be invariant under (\ref{psitransf}) as, at least in the power-counting renormalizable case, the Higgs mechanism cannot modify that term. Since the generators satisfy (\ref{chirality}), we have
\be \Gamma^{\mu}=c^{\mu}_{\,\,\,\nu}\gamma^{\nu} +d^{\mu}_{\,\,\,\nu}\gamma_5 \gamma^{\nu}. \label{Gansatz}\ee
We observe that (\ref{Gansatz}) is also the most general form of $\Gamma^{\mu}$ compatible with the SM gauge group \cite{Colladay:1998fq}. In Ref. \cite{Arias:2007xt} the authors assumed $d^{\mu}_{\,\,\,\nu}$ to be proportional to $c^{\mu}_{\,\,\,\nu}$ and studied the singlet anomaly for a massless fermion in a particular regularization. In Ref. \cite{Coleman:1998ti} instead the rotational invariant case has been considered for the gauge anomalies. 
 Here, we do not require these additional properties. Our results are however consistent with those of Refs. \cite{Coleman:1998ti,Arias:2007xt}.

To study anomalies we introduce in the usual way the functional $W[A]$:
\be e^{i\,W[A]} \equiv \int \delta \psi \delta \overline{\psi} \,e^{ i\, S[A] },\ee 	
with the normalization of the fermion measure chosen in a way that $\exp\left(i\,W[0]\right)=1$. As usual the absence of anomalies corresponds  to the gauge invariance of $W[A]$ under (\ref{Atransf}):
\be \delta W[A] = 0 \quad + \,\,\,\mbox{$M$-terms}, \quad \mbox{(in the absence of anomalies)},\label{no-anomal}\ee
where  $M$-terms represent the non invariance of $W[A]$ due to non gauge invariant terms in $-\overline{\psi}M\psi$, if any. Condition (\ref{no-anomal}) is equivalent to the Ward identities (WIs) for the n-point Green functions
\be \langle J^{\mu_1}_{b_1}(x_1) ... J^{\mu_n}_{b_n}(x_n)\rangle=\int \delta \psi \delta \overline{\psi}\,e^{ i\, S[A=0] }J^{\mu_1}_{b_1}(x_1) ... J^{\mu_n}_{b_n}(x_n), \label{Green}\ee
like in the Lorentz invariant case. However, now we have to change the definition of the currents according to our classical action:
\be J^{\mu}_b \equiv \overline{\psi}\,  \Gamma^{\mu} T_b \psi, \quad \mbox{with} \quad T_b \equiv t^b_L P_L  + t^b_R P_R.  \ee
In the following we will also use the left-right basis and the vector-axial basis for the currents $J^{\mu}_b = J^{\mu}_{bL} + J^{\mu}_{bR} = j^{\mu}_b + j^{\mu}_{b5}$, where
\be J^{\mu}_{bL}\equiv \overline{\psi} \Gamma^{\mu}t^b_L P_L \psi,\quad   J^{\mu}_{bR}\equiv \overline{\psi} \Gamma^{\mu}t^b_R P_R \psi,\quad  j^{\mu}_b\equiv  \overline{\psi} \Gamma^{\mu}t^b \psi,\quad j^{\mu}_{b5}\equiv  \overline{\psi} \Gamma^{\mu}\gamma_5t^b_5  \psi,\ee
with $t^b\equiv (t^b_L + t^b_R)/2$ and $t^b_5\equiv (t^b_L - t^b_R)/2$.

The WIs for (\ref{Green}) can be derived from the functional integral by assuming the invariance of the fermion measure. As usual the anomalies can be thought as a non trivial Jacobian associate with a transformation of the form (\ref{psitransf}) and therefore in perturbation theory corresponds to a one-loop effect \cite{Bilal:2008qx}. For this reason  we shall only consider one-loop contributions in this paper.

\section{Triangle graphs} \label{fermionredef}

Here we show that, in particular regularizations, the anomalous part of the 3-point functions is insensitive to the Lorentz violating parameters for the fermion model previously defined. 

 Let us start by considering the anomalous part of $\delta W[A]$ in the Lorentz invariant case ($\Gamma^{\mu}=\gamma^{\mu}$ and $M=m$). This object may be written \cite{Zumino:1983rz} as follows:
\be \delta W[A]_{\,anom} = \frac{1}{48\pi^2}  \mbox{Tr}\int d^4x  \,\epsilon^{\mu \nu \lambda \rho} \Omega_L \,\partial_{\mu} \left(2 A_{\nu}^{L} \partial_{\lambda}A_{\rho}^L -i A_{\nu}^LA_{\lambda}^LA_{\rho}^L\right)- (L\rightarrow R), \label{LIdeltaW}\ee
where we have defined $$\Omega_{L(R)} \equiv \Omega^b t^b_{L(R)}, \qquad   A_{\mu}^{L(R)} \equiv A_{\mu}^b t^b_{L(R)},$$ and $\epsilon^{\kappa \nu \lambda \rho}$ is the totally antisymmetric quantity with $\epsilon^{0123}=1$. The part in (\ref{LIdeltaW}) that is quadratic in $A^{L(R)}_{\mu}$ can be computed by evaluating the 3-point functions \cite{Bell:1969ts} in a particular regularization, whereas the remaining terms can be obtained by using the Wess-Zumino consistency condition \cite{Wess:1971yu} (for a review see e.g. \cite{Treiman:1986ep, Weinberg:1996kr}). A property of (\ref{LIdeltaW}), which will be used in the following, is the invariance under general coordinate transformations that holds separately for the left-handed and right-handed parts.

In this section we study the part in $\delta W[A]_{\,anom}$, which is quadratic in $A_{\mu}^b$, in the presence of the Lorentz violating parameters, by using an explicit one-loop computation of the 3-point functions. In order to do so we make use of the free ($A_{\mu}^b=0$) fermion propagator. This object may be written (for small Lorentz violating parameters) as follows:
\be \Sigma(x) \equiv \int_{C_F} \frac{d^4p}{(2\pi)^4} \frac{i}{\Gamma^{\mu} p_{\mu} - M} e^{-i\,px}, \label{propagator}\ee
where $C_F$ is a generalized Feynman contour in the $p_0$ space \cite{Colladay:1996iz,Kostelecky:2000mm}.
However, for our purposes it is sufficient to consider (\ref{propagator}) just as a compact way to write the perturbative expansion w.r.t.  
\be \delta \Gamma^{\mu} \equiv \Gamma^{\mu} -\gamma^{\mu}\quad \mbox{and} \quad \delta M \equiv M -m,\ee
like in Ref. \cite{Kostelecky:2001jc}. The propagator in the momentum space $\tilde{\Sigma}(p)$ 
is therefore:
\bea \tilde{\Sigma}(p) &=& \frac{i}{ \displaystyle{\not} p -m+i\epsilon}+\frac{i}{ \displaystyle{\not} p -m+i\epsilon}i \delta\Gamma^{\mu}p_{\mu}\frac{i}{ \displaystyle{\not} p -m+i\epsilon}\nonumber \\&&+\frac{i}{ \displaystyle{\not} p -m+i\epsilon}\left(-i \delta M\right)\frac{i}{ \displaystyle{\not} p -m+i\epsilon}+...,\label{expansion}\eea 
where $\displaystyle{\not} p \equiv \gamma^{\mu} p_{\mu}$. The second and third terms in the right hand side of (\ref{expansion}) are the linear contributions of $\delta \Gamma^{\mu}$ and $\delta M $ to $\tilde{\Sigma}(p)$; the dots represent instead all the higher order contributions that are given by the formal perturbative expansion of $i/(\Gamma^{\mu} p_{\mu} - M)$ w.r.t. $\delta \Gamma^{\mu}$ and $\delta M $. Finally, $i/(\displaystyle{\not} p -m+i\epsilon)$ is the standard fermion propagator with the usual $i\epsilon$-prescription:
$i/( \displaystyle{\not} p -m+i\epsilon)\equiv i ( \displaystyle{\not} p +m)/(p^2-m^2+i\epsilon)$.

\subsection{Massless case}\label{massless}
We start by considering case $m+H^{\mu \nu} \sigma_{\mu \nu}/2=0$. We shall deal with $m+H^{\mu \nu} \sigma_{\mu \nu}/2\neq 0$ in the next subsection. We refer to this case as the massless case because, as we shall see, the triangle graphs in a sense can be reduced to the massless Lorentz invariant case.

\begin{figure}
\centering
\epsfig{file=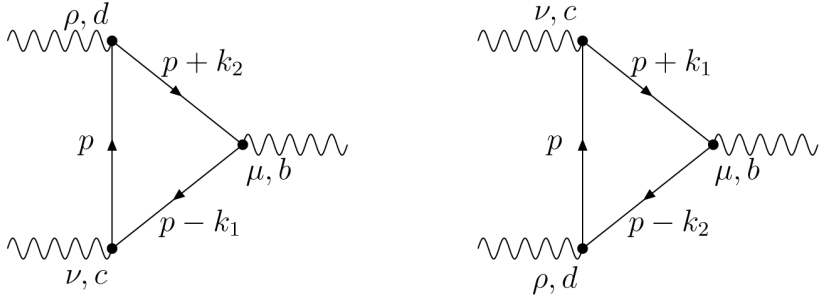,width=0.7\linewidth,clip=}
\caption{\small The triangle graphs in the momentum space with a particular assignment of the internal momenta.}\label{graphs}
\end{figure}

We write the 3-point functions as follows:
\be \langle J^{\mu}_b(x) J^{\nu}_c(y)J^{\rho}_d(z)\rangle=\langle J^{\mu}_{Lb}(x) J^{\nu}_{c}(y)J^{\rho}_{d}(z)\rangle +\langle J^{\mu}_{Rb}(x) J^{\nu}_{c}(y)J^{\rho}_{d}(z)\rangle,\ee 
where we have expanded the first current in the left-right basis. In the case at hand the left-handed part can be written as
\bea && \langle J^{\mu}_{Lb}(x) J^{\nu}_{c}(y)J^{\rho}_{d}(z)\rangle =\langle J^{\mu}_{Lb}(x) J^{\nu}_{Lc}(y)J^{\rho}_{Ld}(z)\rangle \nonumber \\ &&=\frac{i}{(2\pi)^{12}} \int d^4k_1 d^4 k_2 e^{i\left[\left(k_1+k_2\right)x-k_1 y - k_2 z\right]}\int d^4 p\,\,\mbox{Tr}\left\{ t^b_L t^c_L t^d_L  \right.\nonumber \\ \hspace{-0.5cm}&&\times \frac{1}{(p-k_1+\alpha)_- -\displaystyle{\not} a +\displaystyle{\not} b} \hspace{0.1cm}\Gamma_L^{\nu}\frac{1}{(p+\alpha)_- -\displaystyle{\not} a +\displaystyle{\not} b} \hspace{0.1cm}\Gamma_L^{\rho}\frac{1}{(p+k_2 +\alpha)_- -\displaystyle{\not} a +\displaystyle{\not} b} \hspace{0.1cm}\Gamma_L^{\mu}P_L \nonumber \\&&+\,\,t^b_L t^d_L t^c_L \nonumber \\  &&\hspace{-0.5cm}\left.\times\frac{1}{(p-k_2-\alpha)_- +\displaystyle{\not} a -\displaystyle{\not} b} \hspace{0.1cm}\Gamma_L^{\rho}\frac{1}{(p-\alpha)_- +\displaystyle{\not} a -\displaystyle{\not} b} \hspace{0.1cm}\Gamma_L^{\nu}\frac{1}{(p+k_1 -\alpha)_- +\displaystyle{\not} a -\displaystyle{\not} b}\hspace{0.1cm} \Gamma_L^{\mu}P_L \right\},\label{divergence} \eea
the right-handed part can be obtained from (\ref{divergence}) by means of the replacements $L\rightarrow R, b_{\mu} \rightarrow -b_{\mu}$ and $\alpha_{\mu} \rightarrow \delta_{\mu}$ and for an arbitrary momentum $p_{\mu}$, we have defined $p_{\mp} \equiv \Gamma^{\mu}_{L(R)} p_{\mu}$, with $\Gamma^{\mu}_{L(R)}\equiv \left(c^{\mu}_{\,\,\,\nu} \mp d^{\mu}_{\,\,\,\nu}\right) \gamma^{\nu}$. 
To derive Eq. (\ref{divergence}) we have used the following identities (for an arbitrary momentum $p_{\mu}$):
\be \frac{1}{\Gamma^{\mu}p_{\mu} -\displaystyle{\not} a-\gamma_5 \displaystyle{\not} b} \gamma^{\nu}P_{L(R)} =\frac{1}{p_{\mp} -\displaystyle{\not} a\pm\displaystyle{\not} b} \gamma^{\nu
} P_{L(R)}\quad  \mbox{and} \quad \Gamma^{\mu} P_{L(R)} = \Gamma_{L(R)}^{\mu} P_{L(R)}. \label{defs}\ee 
Also we have introduced the constant four-vectors  $\alpha_{\mu}$ and $\delta_{\mu}$, which parametrize the ambiguity of the  divergent integrals  due to different labeling of the integration variables, just like in Lorentz invariant theories (see e.g. \cite{Treiman:1986ep,Weinberg:1996kr}). The constant four-vector $\alpha_{\mu}$ has been introduced by performing the following replacements in the assignment displayed in Figure \ref{graphs}: $p_{\mu}\rightarrow p_{\mu}+\alpha_{\mu}$ and $p_{\mu}\rightarrow p_{\mu}+\beta_{\mu}$ (where $\beta_{\mu}$ is the solution to $\beta_-= -\alpha_-+2\displaystyle{\not} a -2\displaystyle{\not} b$) respectively in the graph on the left and on the right in Figure \ref{graphs}. 

We observe that the generalized Dirac matrices $\Gamma^{\mu}=c^{\mu}_{\,\,\,\nu}\gamma^{\nu} + d^{\mu}_{\,\,\,\nu}\gamma_5 \gamma^{\nu}$ simply reduce to linear combinations of the ordinary Dirac matrices in the left-handed and right-handed parts, respectively $\Gamma_L^{\mu}=l^{\mu}_{\,\,\,\nu} \gamma^{\nu}$ and $ \Gamma_R^{\mu}=r^{\mu}_{\,\,\,\nu} \gamma^{\nu}$, where 
\be l^{\mu}_{\,\,\,\nu}\equiv c^{\mu}_{\,\,\,\nu}-d^{\mu}_{\,\,\,\nu}\quad \mbox{and} \quad r^{\mu}_{\,\,\,\nu}\equiv c^{\mu}_{\,\,\,\nu}+d^{\mu}_{\,\,\,\nu}. \ee
Although these linear combinations are in general different (as in general $l^{\mu}_{\,\,\,\nu} \neq r^{\mu}_{\,\,\,\nu}$), in this subsection we will show that the anomalous terms in the WIs are independent of both  $l^{\mu}_{\,\,\,\nu}-\delta^{\mu}_{\nu}$ and $r^{\mu}_{\,\,\,\nu}-\delta^{\mu}_{\nu}$ and therefore of both $c^{\mu}_{\,\,\,\nu}-\delta^{\mu}_{\nu}$ and $d^{\mu}_{\,\,\,\nu}$ (in a certain regularization). As we shall see, this is related to the fact that the left-handed and the right-handed parts of $\delta W[A]_{anom}$ in (\ref{LIdeltaW}) are separately invariant under general coordinate transformations. In this and the next subsections we shall also demonstrate the $M$-independence of the anomaly.

Since the integral w.r.t. $p$ in (\ref{divergence}) has a linear superficial  divergence, we regularize it by integrating only over a finite domain $D$, whose volume will go to infinity at the end of the calculations; for the moment we do not fix the shape of $D$. As we shall see, after a redefinition of the integration variables, below we shall identify $D$ with the domain inside a 3-sphere with a very large radius $P$.

We focus now on the left-handed 3-point functions in (\ref{divergence}) and we will explain afterwards  how to deal with the right-handed ones. A formal functional integral derivation (in which one assumes the invariance of the fermion measure under (\ref{psitransf})) leads to the following  WIs for the one-loop 3-point functions:
\bea i\frac{\partial}{\partial x^{\mu}}\langle J^{\mu}_{Lb}(x) J^{\nu}_{Lc}(y)J^{\rho}_{Ld}(z)\rangle =f^{cb h} \delta(x-y) \langle J^{\nu}_{Lh}(y) J^{\rho}_{Ld}(z)\rangle \nonumber \\+ f^{db h} \delta(x-z) \langle J^{\nu}_{Lc}(y) J^{\rho}_{Lh}(z)\rangle. \label{WIL}\eea 
By performing a redefinition of the momentum shift $\alpha_{\mu}$ in (\ref{divergence}) such that
\be \alpha_- -\displaystyle{\not} a +\displaystyle{\not} b =\displaystyle{\not} \alpha' \label{shift-redef}\ee
and  using the identities
\bea \left(k_1+k_2\right)_- \frac{1}{\left(p-k_1\right)_- + \displaystyle{\not} \alpha'}&=& -1 +\left[ \left(p+k_2\right)_-+\displaystyle{\not} \alpha'\right] \frac{1}{\left(p-k_1\right)_- + \displaystyle{\not} \alpha'},\nonumber \\\left(k_1+k_2\right)_- \frac{1}{\left(p-k_2\right)_- - \displaystyle{\not} \alpha'}&=& -1 +\left[ \left(p+k_1\right)_--\displaystyle{\not} \alpha'\right] \frac{1}{\left(p-k_2\right)_- - \displaystyle{\not} \alpha'},
\eea
we obtain the following divergence of the 3-point functions
\bea \frac{\partial}{\partial x^{\mu}}\langle J^{\mu}_{Lb}(x) J^{\nu}_{Lc}(y)J^{\rho}_{Ld}(z)\rangle = -\int \frac{d^4k_1}{(2\pi)^4} \frac{d^4 k_2}{(2\pi)^4} e^{i\left[\left(k_1+k_2\right)x-k_1 y - k_2 z\right]}\int \frac{d^4 p}{(2\pi)^4}\,\,\nonumber \\ \times\mbox{Tr}\left\{ t^b_L t^c_Lt^d_L  \left[- \Gamma^{\nu}_L \frac{1}{p_- + \displaystyle{\not} \alpha'}\Gamma^{\rho}_L P_L\frac{1}{\left(p+k_2\right)_- + \displaystyle{\not} \alpha'} +\frac{1}{\left(p-k_1\right)_- + \displaystyle{\not} \alpha'}\Gamma^{\nu}_L  \frac{1}{p_- + \displaystyle{\not} \alpha'}\Gamma^{\rho}_L P_L\right]\right.\nonumber \\\left.+ t^b_L t^d_L t^c_L \left[- \Gamma^{\rho}_L \frac{1}{p_- - \displaystyle{\not} \alpha'}\Gamma^{\nu}_L P_L\frac{1}{\left(p+k_1\right)_- - \displaystyle{\not} \alpha'} +\frac{1}{\left(p-k_2\right)_- - \displaystyle{\not} \alpha'}\Gamma^{\rho}_L  \frac{1}{p_- - \displaystyle{\not} \alpha'}\Gamma^{\nu}_L P_L\right]\right\}.\label{divergenceL}\eea 

On the other hand, the 2-point functions in the right-hand side of (\ref{WIL}) have the following one-loop representation:
\bea &&\langle J^{\nu}_{Lh}(y) J^{\rho}_{Ld}(z)\rangle =\nonumber \\ &&\hspace{-0.5cm}N\delta^{hd}\hspace{-0.2cm}\int \frac{d^4k_2}{(2\pi)^4} e^{ik_2(y-z)}\mbox{Tr}\hspace{-0.1cm}\int\frac{d^4 p}{(2\pi)^4}\frac{1}{\left(p-k_2/2+\omega \right)_- -\displaystyle{\not} a +\displaystyle{\not} b}\Gamma^{\rho}_LP_L \frac{1}{\left(p+k_2/2+\omega \right)_- -\displaystyle{\not} a +\displaystyle{\not} b}\Gamma^{\nu}_L \nonumber \\&& =N\delta^{hd}\hspace{-0.2cm}\int \frac{d^4k_2}{(2\pi)^4} e^{ik_2(y-z)}\mbox{Tr}\hspace{-0.1cm}\int\frac{d^4 p}{(2\pi)^4}\frac{1}{\left(p-k_2/2 \right)_- +\displaystyle{\not} \omega'}\Gamma^{\rho}_LP_L \frac{1}{\left(p+k_2/2 \right)_- +\displaystyle{\not} \omega'}\Gamma^{\nu}_L, \label{2-pointL}\eea
where $N$ is defined by Tr$\{t^h_Lt^d_L\} = N \delta^{hd}$ and we introduced another constant four-vector $\omega_{\mu}$, which keeps track of the ambiguity due to the divergence of this integral; also in the last line of (\ref{2-pointL}) we redefined such four-vector like in (\ref{shift-redef}): $\omega_- -\displaystyle{\not} a +\displaystyle{\not} b =\displaystyle{\not} \omega'$. The superficial quadratic divergence in (\ref{2-pointL}) has been regularized by performing the p-integration only over a finite domain $D$, like in (\ref{divergence}). We observe that the redefinition in (\ref{shift-redef}) and $\omega_- -\displaystyle{\not} a +\displaystyle{\not} b =\displaystyle{\not} \omega'$ reabsorb the Lorentz violating parameters $a_{\mu}$ and $b_{\mu}$, but do not eliminate the dependence on $\delta \Gamma^{\mu}$. In order to deal with the latter, we perform the following changes of variable in integrals (\ref{divergenceL}) and (\ref{2-pointL}):
\be l^{\mu}_{\,\,\,\nu}p_{\mu}=p'_{\nu}, \quad l^{\mu}_{\,\,\,\nu} k_{1\mu}=k'_{1\nu}, \quad l^{\mu}_{\,\,\,\nu}k_{2\mu}=k'_{2\nu},  \label{redef}
\ee
which eliminate the dependence on the Lorentz violating parameters in the denominators of the propagator, e.g. $p_- + \displaystyle{\not} \alpha'=\displaystyle{\not} p\,' + \displaystyle{\not} \alpha' $, but lead to non-trivial Jacobians   in (\ref{divergenceL}) and (\ref{2-pointL}): respectively $l^{-3}$ and $l^{-2}$, where $l\equiv |$det$\{l^{\mu}_{\,\,\,\nu}\}|$. The redefinitions in (\ref{redef}) are allowed because the smallness of the Lorentz violating parameters ensures that the $ l^{\mu}_{\,\,\,\nu}$ matrix is invertible. Of course redefinition $l^{\mu}_{\,\,\,\nu}p_{\mu}=p'_{\nu}$ in (\ref{redef}) changes the domain $D$ to some new domain $D'$, which however remains with finite volume (as long as the cut-off is not removed) and shape still not fixed. 
It is now easy to show 
\bea \frac{\partial}{\partial x^{\mu}}\langle J^{\mu}_{Lb}(x) J^{\nu}_{Lc}(y)J^{\rho}_{Ld}(z)\rangle =  -\,l^{-3}\,l^{\nu}_{\,\,\,\tau}l^{\rho}_{\,\,\,\sigma}\int \frac{d^4k_1}{(2\pi)^4} \frac{d^4 k_2}{(2\pi)^4} e^{i\left[\left(k_1+k_2\right)x'-k_1 y' - k_2 z'\right]}\int \frac{d^4 p}{(2\pi)^4}\,\,\nonumber \\ \times\mbox{Tr}\left\{ t^b_L t^c_Lt^d_L  \left[- \gamma^{\tau} \frac{1}{\displaystyle{\not} p + \displaystyle{\not} \alpha'}\gamma^{\sigma} P_L\frac{1}{\displaystyle{\not} p+\displaystyle{\not} k_2 + \displaystyle{\not} \alpha'} +\frac{1}{\displaystyle{\not} p-\displaystyle{\not} k_1 + \displaystyle{\not} \alpha'}\gamma^{\tau}  \frac{1}{\displaystyle{\not} p + \displaystyle{\not} \alpha'}\gamma^{\sigma} P_L\right]\right.\nonumber \\\left.+ t^b_L t^d_L t^c_L \left[- \gamma^{\sigma} \frac{1}{\displaystyle{\not} p - \displaystyle{\not} \alpha'}\gamma^{\tau} P_L\frac{1}{\displaystyle{\not} p+\displaystyle{\not} k_1 - \displaystyle{\not} \alpha'} +\frac{1}{\displaystyle{\not}p-\displaystyle{\not} k_2 - \displaystyle{\not} \alpha'}\gamma^{\sigma}  \frac{1}{\displaystyle{\not}p - \displaystyle{\not} \alpha'}\gamma^{\tau} P_L\right]\right\},\label{step}\eea 
where  $l^{\nu}_{\,\,\,\mu}x'^{\mu}\equiv x^{\nu}$,  $l^{\nu}_{\,\,\,\mu}y'^{\mu}=y^{\nu}$ and $l^{\nu}_{\,\,\,\mu}z'^{\mu}=z^{\nu}$. In other words we can write
\be \frac{\partial}{\partial x^{\mu}}\langle J^{\mu}_{Lb}(x) J^{\nu}_{Lc}(y)J^{\rho}_{Ld}(z)\rangle = l^{-3}\,l^{\nu}_{\,\,\,\tau}l^{\rho}_{\,\,\,\sigma}\frac{\partial}{\partial x'^{\mu}}\langle J^{\mu}_{Lb}(x') J^{\tau}_{Lc}(y')J^{\sigma}_{Ld}(z')\rangle'_{LI},\label{LI-LV}
\ee
 where the label $LI$ indicates the Lorentz invariant case ($\delta \Gamma^{\mu}=0, \delta M=0$). The prime outside the angle bracket in the right hand side of (\ref{LI-LV}) reminds that those Green functions are computed in the redefined constant four vector $\alpha'_{\mu}$, as it is clear in (\ref{step}).
Likewise one can show
\be \langle J^{\nu}_{Lh}(y) J^{\rho}_{Ld}(z)\rangle = l^{-2}\,l^{\nu}_{\,\,\,\tau}l^{\rho}_{\,\,\,\sigma}\langle J^{\tau}_{Lh}(y') J^{\sigma}_{Ld}(z')\rangle'_{LI}.\label{LI-LV2point}\ee 
Again the prime outside the angle bracket in the right hand side of (\ref{LI-LV2point}) indicates that those Green functions are computed in the redefined constant four vector $\omega'_{\mu}$. Since we assumed $m+H^{\mu \nu} \sigma_{\mu \nu}/2=0$,  the $LI$ Green functions in (\ref{LI-LV}) and (\ref{LI-LV2point}) are in the massless case ($m=0$). It is now clear that, if the 3-point (and 2-point) functions satisfy the WIs in the Lorentz invariant case, so they do in the Lorentz violating case\footnote{The difference in the power of $l$ in (\ref{LI-LV}) and (\ref{LI-LV2point}) is compensated by the transformation rule $\delta(x-y)=\delta(x'-y')/l$ of the Dirac delta appearing in the right-hand side of (\ref{WIL}).}. It is well-known that this happens\footnote{This is indeed true for the choice of the constant four-vector $\beta_{\mu}$ discussed below Eq. (\ref{defs}) and for some value of the four-vector $\omega'_{\mu}$.} for the antisymmetric part of the 3-point functions \cite{Bardeen:1969md}, that is (\ref{divergence}), where $t^b_L t^c_Lt^d_L $ and $t^b_L t^d_L t^c_L $ 
 are replaced with  $t^b_L [t^c_L,t^d_L]/2 $ and $t^b_L [t^d_L ,t^c_L]/2 $ respectively. Therefore, like in the Lorentz invariant case, the anomaly is contained only in the symmetric part, namely (\ref{divergenceL}) with $t^b_L t^c_Lt^d_L $ and $t^b_L t^d_L t^c_L $ replaced with $t^b_L \{t^c_L,t^d_L\}/2 $ and $t^b_L \{t^d_L, t^c_L\}/2 $ respectively. So we focus on the symmetric part of Eq. (\ref{LI-LV}):
\be \frac{\partial}{\partial x^{\mu}}\langle J^{\mu}_{Lb}(x) J^{\nu}_{Lc}(y)J^{\rho}_{Ld}(z)\rangle_{anom} = l^{-3}\,l^{\nu}_{\,\,\,\tau}l^{\rho}_{\,\,\,\sigma}\frac{\partial}{\partial x'^{\mu}}\langle J^{\mu}_{Lb}(x') J^{\tau}_{Lc}(y')J^{\sigma}_{Ld}(z')\rangle'_{LI\,anom},\label{LV-LIanom}
\ee
where the label $anom$ indicates the anomalous part or equivalently the symmetric part. Since $l^{-3}\,l^{\nu}_{\,\,\,\tau}l^{\rho}_{\,\,\,\sigma}$ is in general not equal to $\delta^{\nu}_{\tau} \delta^{\rho}_{\sigma}$, the anomalous part looks to depend on the Lorentz violating parameters. However, we expect the general coordinate invariance of (\ref{LIdeltaW}) to correspond to a symmetry of $\langle J^{\mu}_{Lb}(x) J^{\nu}_{Lc}(y)J^{\rho}_{Ld}(z)\rangle_{LI\,anom}$, which indeed satisfies (see the explanation below)
\be \frac{\partial}{\partial x^{\mu}}\langle J^{\mu}_{Lb}(x) J^{\nu}_{Lc}(y)J^{\rho}_{Ld}(z)\rangle_{LI\,anom}= l^{-3}\,l^{\nu}_{\,\,\,\tau}l^{\rho}_{\,\,\,\sigma}\frac{\partial}{\partial x'^{\mu}}\langle J^{\mu}_{Lb}(x') J^{\tau}_{Lc}(y')J^{\sigma}_{Ld}(z')\rangle''_{LI\,anom}, \label{property}\ee
for any invertible $l^{\mu}_{\,\,\,\nu}$ matrix and the double prime outside the angle bracket indicates that those Green functions are computed in the constant four-vector   
\be \alpha''_{\mu}\equiv l^{\nu}_{\,\,\,\mu}\alpha_{\nu}.\label{Red2}\ee
 The latter feature can be shown by using the explicit expression of $\langle J^{\mu}_{Lb}(x) J^{\nu}_{Lc}(y)J^{\rho}_{Ld}(z)\rangle_{LI\,anom}$ given in the literature (see e.g. \cite{Weinberg:1996kr}):
\bea  &&\frac{\partial}{\partial x^{\mu}}\langle J^{\mu}_{Lb}(x) J^{\nu}_{Lc}(y)J^{\rho}_{Ld}(z)\rangle_{LI\,anom}=\nonumber \\ &&-\frac{\pi^2}{(2 \pi)^{12}}\mbox{Tr}\left(t^b_L \{t^c_L,t^d_L\}\right) \int d^4 k_1 d^4 k_2 e^{i\left[\left(k_1+k_2\right)x-k_1 y - k_2 z\right]} \epsilon^{\kappa \nu \lambda \rho} \alpha_{\kappa} (k_1+k_2)_{\lambda}.\label{explicit-anom}\eea
 The freedom to choose $\alpha_{\mu}$ corresponds to a freedom to move the anomaly from one current to another, but no choice of the momentum labels can remove the anomaly altogether; in the Lorentz invariant case the choice $\alpha_{\mu}=(k_{1\mu}-k_{2\mu})/3$ corresponds to the anomaly symmetrically distributed on the three currents and to the form of $\delta W[A]_{anom}$ given in (\ref{LIdeltaW}).

 Expression (\ref{explicit-anom}) has been obtained in \cite{Weinberg:1996kr} by performing a Wick rotation, evaluating the $p$-integration over the domain inside a 3-sphere with radius $P$  and then taking the limit $P\rightarrow \infty$. We observe that, our initial integration domain $D$, since with arbitrary shape, is compatible with this set-up. We also note that result (\ref{explicit-anom}) has been derived in a particular regularization.

By using (\ref{explicit-anom}), the right hand side of (\ref{property}) is
\bea l^{-3}\,l^{\nu}_{\,\,\,\tau}l^{\rho}_{\,\,\,\sigma}\frac{\partial}{\partial x'^{\mu}}\langle J^{\mu}_{Lb}(x') J^{\tau}_{Lc}(y')J^{\sigma}_{Ld}(z')\rangle''_{LI\,anom} =-\frac{\pi^2}{(2 \pi)^{12}}\mbox{Tr}\left(t^b_L \{t^c_L,t^d_L\}\right)\nonumber \\ \times \, l^{-2} \int d^4 k_1 d^4 k_2 e^{i\left[\left(k_1+k_2\right)x'-k_1 y' - k_2 z'\right]} l^{\nu}_{\,\,\,\tau}l^{\rho}_{\,\,\,\sigma}l^{-1}\epsilon^{\kappa \tau \lambda \sigma} \alpha''_{\kappa} (k_1+k_2)_{\lambda}\eea
and property (\ref{property}) can be checked by performing the changes of integration variables $k_{1(2)\mu}=l^{\nu}_{\,\,\,\mu}k'_{1(2)\nu}$ and remembering the definition $\alpha''_{\mu}\equiv l^{\nu}_{\,\,\,\mu}\alpha_{\nu}$.
 By combining (\ref{property}) with (\ref{LV-LIanom}) it follows that the anomalous part of the 3-point functions coincides with its Lorentz invariant limit up to an unimportant redefinition of the constant four-vector $\alpha_{\mu}$.

We observe that one can prove  that $\frac{\partial}{\partial x^{\mu}}\langle J^{\mu}_{Rb}(x) J^{\nu}_{Rc}(y)J^{\rho}_{Rd}(z)\rangle_{anom}$ is independent of the Lorentz violating parameters (for a certain regularization) in a similar way: as we have commented before, although  in the right-handed part of the 3-point functions the matrix $l^{\mu}_{\,\,\,\nu}\equiv c^{\mu}_{\,\,\,\nu}-d^{\mu}_{\,\,\,\nu}$ is replaced with $r^{\mu}_{\,\,\,\nu} \equiv c^{\mu}_{\,\,\,\nu}+d^{\mu}_{\,\,\,\nu}$, the left-handed and the right-handed parts of (\ref{LIdeltaW}) are separately invariant under general coordinate transformations and so the argument presented here is applicable also to case $r^{\mu}_{\,\,\,\nu}\neq l^{\mu}_{\,\,\,\nu}$. Indeed, in the right-handed part one obtains relations identical to (\ref{LV-LIanom}) and (\ref{property}), apart from the fact that the left-handed quantities are replaced by the right-handed ones (e.g. $l^{\mu}_{\,\,\,\nu}$ replaced with $r^{\mu}_{\,\,\,\nu}$). We would like to stress that, as a consequence of this fact, the anomalous terms in the WIs turn out to be independent of both $c^{\mu}_{\,\,\,\nu}-\delta^{\mu}_{\nu}$ and $d^{\mu}_{\,\,\,\nu}$, without any further assumption on $d^{\mu}_{\,\,\,\nu}$. Moreover, this result is valid also if the fermion representation is made by more than one irreps because (\ref{LIdeltaW}) can be decomposed as a sum over the irreps, which are separately invariant under general coordinate transformations.

We finally note that in our reasoning we explicitly used that (\ref{property}) is satisfied by $\langle J^{\mu}_{Lb}(x) J^{\nu}_{Lc}(y)J^{\rho}_{Ld}(z)\rangle_{LI\,anom}$  and that the antisymmetric part does not violate the WIs in the Lorentz invariant case. This follows from the explicit expression of   $\langle J^{\mu}_{Lb}(x) J^{\nu}_{Lc}(y)J^{\rho}_{Ld}(z)\rangle_{LI}$  derived in the literature by using Lorentz invariant regulators or anyway particular Lorentz violating regulators. Therefore, the argument presented here shows that there are regulators for which the 3-point functions are insensitive to the Lorentz invariant parameters. A regulator independent proof of the invariance of the anomaly cancellation conditions under Lorentz violations (which also takes into account general Lorentz violating regulators) will be provided in Section  \ref{regulator-indep}.

\subsection{Massive case}

We now consider  case $M_1\equiv m+H^{\mu \nu} \sigma_{\mu \nu}/2\neq 0$ and show that the anomalous part of the 3-point functions are independent of $M_1$ even if the Lorentz violating term $H^{\mu \nu} \sigma_{\mu \nu}/2$ is included.

Here it is not convenient to use the left-right basis for the currents as now the fermion propagator contains a part that commutes with $\gamma_5$ and a part that anticommutes with $\gamma_5$. We choose the vector-axial basis to write the 3-point functions and we focus on the vector-vector-axial functions
\be \langle j^{\mu}_b(x) j^{\nu}_c(y)j^{\rho}_{5d}(z)\rangle, \ee
as other types of Green functions, like the axial-axial-axial functions, can be studied in a similar way. 

  The vector WIs (VWIs) and the axial WIs (AWIs) are respectively (assuming the invariance of the fermion measure under (\ref{psitransf}))
\bea \frac{\partial}{\partial x^{\mu}}\langle j^{\mu}_b(x) j^{\nu}_c(y)j^{\rho}_{5d}(z)\rangle = \delta(x-y) \langle \overline{\psi}(y)[t^b,t^c]\Gamma^{\nu}\psi(y)\,j^{\rho}_{5d}(z)\rangle\nonumber \\+\delta(x-z) \langle j^{\nu}_c(y)\, \overline{\psi}(z)[t^b,t^d_5]\Gamma^{\rho}\gamma_5\psi(z)\rangle\eea
and 
\bea \frac{\partial}{\partial z^{\rho}}\langle j^{\mu}_b(x) j^{\nu}_c(y)j^{\rho}_{5d}(z)\rangle  = 2i\langle j^{\mu}_b(x) j^{\nu}_c(y) P^d(z)\rangle\nonumber \\+\delta(x-z) \langle \overline{\psi}(x)[t^d_5,t^b]\Gamma^{\mu}\gamma_5\psi(x)\,j^{\nu}_{c}(y)\rangle+\delta(y-z) \langle j^{\mu}_b(x)\, \overline{\psi}(y)[t^d_5,t^c]\Gamma^{\nu}\gamma_5\psi(y)\rangle,\eea
where $P^b\equiv \overline{\psi}\{\gamma_5,M\}t^b_5 \psi/2$. 

Now let us take for simplicity $\Gamma^{\mu}=\gamma^{\mu}$ and $a_{\mu}=b_{\mu}=0$, but keep $M_1\neq 0$. It will be straightforward to treat the general case by combining the results of the previous subsection with those that we present here. The one-loop vector-vector-axial function is
\bea && \langle J^{\mu}_{b}(x) J^{\nu}_{c}(y)J^{\rho}_{5d}(z)\rangle  =\frac{i}{(2\pi)^{12}} \int d^4k_1 d^4 k_2 e^{i\left[\left(k_1+k_2\right)x-k_1 y - k_2 z\right]}\nonumber \\ &&\times \int d^4 p\,\,\mbox{Tr}\left\{ t^b t^c t^d_5  \, \frac{1}{\displaystyle{\not}p-\displaystyle{\not}k_1+\displaystyle{\not}\alpha -M_1} \hspace{0.1cm}\gamma^{\nu}\frac{1}{\displaystyle{\not}p+\displaystyle{\not}\alpha - M_1} \hspace{0.1cm}\gamma^{\rho}\gamma_5\frac{1}{\displaystyle{\not}p+\displaystyle{\not}k_2 +\displaystyle{\not}\alpha -M_1} \hspace{0.1cm}\gamma^{\mu}\right. \nonumber \\&&+\,\,t^b t^d_5 t^c \left.\frac{1}{\displaystyle{\not}p-\displaystyle{\not}k_2-\displaystyle{\not}\alpha -M_1} \hspace{0.1cm}\gamma^{\rho}\gamma_5\frac{1}{\displaystyle{\not}p-\displaystyle{\not}\alpha -M_1} \hspace{0.1cm}\gamma^{\nu}\frac{1}{\displaystyle{\not}p+\displaystyle{\not}k_1 -\displaystyle{\not}\alpha -M_1}\hspace{0.1cm} \gamma^{\mu}\right\}, \label{VVA}\eea
where, for completeness, we have again introduced a constant four-vector $\alpha_{\mu}$ even if, as we shall see, it does not play a crucial role here. This four-vector has been introduced by performing the following replacements in the assignment of Figure \ref{graphs}: $p_{\mu}\rightarrow p_{\mu}+\alpha_{\mu}$ and $p_{\mu}\rightarrow p_{\mu}-\alpha_{\mu}$, respectively in the graph on the left and on the right in Figure \ref{graphs}. Like in Subsection \ref{massless}, the linear superficial divergence of (\ref{VVA}) has been regularized by performing the $p$-integration over a domain $D$ with finite volume, but with shape still not fixed. We shall see that the $M_1$-independence of the anomaly can be showed without choosing a particular shape of $D$. 
 We can now split (\ref{VVA}) in the antisymmetric and symmetric part (respectively $ t^b t^c t^d_5, t^b t^d_5 t^c$ replaced with $ t^b [t^c, t^d_5]/2, t^b [t^d_5, t^c]/2$ and $t^b \{t^c, t^d_5\}/2, t^b \{t^d_5 ,t^c\}/2 $). Again the antisymmetric part does not harbour any anomaly like in the Lorentz invariant case (we shall provide an explanation below). Therefore, we focus on the symmetric part. 

Let us start by considering the VWIs. We have
\bea \frac{\partial}{\partial x^{\mu}}\langle j^{\mu}_b(x) j^{\nu}_c(y)j^{\rho}_{5d}(z)\rangle_{anom}= -\int \frac{d^4k_1}{(2\pi)^4} \frac{d^4 k_2}{(2\pi)^4} e^{i\left[\left(k_1+k_2\right)x-k_1 y - k_2 z\right]}\nonumber \\ \times\int \frac{d^4 p}{(2\pi)^4}\mbox{Tr}\left\{t^b \frac{\{t^c,t^d_5\}}{2}   \left[\frac{1}{\displaystyle{\not} p- \displaystyle{\not}k_1+\displaystyle{\not}\alpha-M_1} \gamma^{\nu}\frac{1}{\displaystyle{\not} p  +\displaystyle{\not}\alpha -M_1}\gamma^{\rho}\gamma_5 \right.\right. \nonumber \\ -\gamma^{\nu}\frac{1}{\displaystyle{\not} p +\displaystyle{\not} \alpha -M_1}\gamma^{\rho}\gamma_5  \frac{1}{\displaystyle{\not} p+ \displaystyle{\not} k_2+\displaystyle{\not} \alpha-M_1} 
  + \frac{1}{\displaystyle{\not} p- \displaystyle{\not} k_2-\displaystyle{\not} \alpha-M_1} \gamma^{\rho}\gamma_5\frac{1}{\displaystyle{\not} p-\displaystyle{\not} \alpha-M_1}\gamma^{\nu} \nonumber \\ \left. \right. -\gamma^{\rho}\gamma_5\frac{1}{\displaystyle{\not} p-\displaystyle{\not} \alpha-M_1}\gamma^{\nu}\frac{1}{\displaystyle{\not} p+ \displaystyle{\not}k_1-\displaystyle{\not}\alpha-M_1}\left]\right\},  \label{anomVVA}\eea
where again the label $anom$ represents the anomalous part or equivalently the symmetric part and we used the identities
\bea \left(\displaystyle{\not} k_1+ \displaystyle{\not}k_2\right) \frac{1}{\displaystyle{\not}p-\displaystyle{\not}k_1+\displaystyle{\not}\alpha -M_1}&=& -1 +\left(\displaystyle{\not} p+\displaystyle{\not}k_2+\displaystyle{\not}\alpha -M_1\right)\frac{1}{\displaystyle{\not}p-\displaystyle{\not}k_1+\displaystyle{\not}\alpha -M_1},\nonumber \\\left(\displaystyle{\not} k_1+ \displaystyle{\not}k_2\right) \frac{1}{\displaystyle{\not}p-\displaystyle{\not}k_2-\displaystyle{\not}\alpha -M_1}&=& -1 +\left(\displaystyle{\not} p+\displaystyle{\not}k_1-\displaystyle{\not}\alpha -M_1\right)\frac{1}{\displaystyle{\not}p-\displaystyle{\not}k_2-\displaystyle{\not}\alpha -M_1}. 
\eea
 We start by considering the first  and fourth terms inside the square brackets of (\ref{anomVVA}). The trace of the integral over $d^4p$ of those terms can be written as follows:
\be I^{\nu \rho}(q) \equiv \int d^4 p \left[h^{\nu \rho}(p+q)- h^{\nu \rho}(p)\right], \label{diff}\ee
where $q_{\mu}\equiv 2\alpha_{\mu} -k_{1\mu}$ and 
\be   h^{\nu \rho}(p)\equiv \mbox{Tr}\left( \gamma^{\rho}\gamma_5\frac{1}{\displaystyle{\not} p-\displaystyle{\not} \alpha-M_1}\gamma^{\nu}\frac{1}{\displaystyle{\not} p+ \displaystyle{\not}k_1-\displaystyle{\not}\alpha-M_1}\right). \label{hmu}\ee
The integral $\int d^4 p \,h^{\nu \rho}(p)$ has a superficial quadratic divergence\footnote{An analysis of (\ref{hmu}) reveals that the divergence is only linear. However, we keep thinking that the divergence is at most quadratic in order for our argument to be applicable to other types of 3-point functions, such as the axial-axial-axial ones.} and therefore we are not allowed to perform a shift in the integration variable and declare that (\ref{diff}) vanishes. This is what leads to the anomaly also in the Lorentz invariant case. However, our aim is to show that the anomalous part of the 3-point functions is independent of $\delta M_1\equiv H^{\mu \nu} \sigma_{\mu \nu}/2$ and to this end we substitute  (\ref{expansion}) into (\ref{hmu}). The terms that are at least quadratic in $\delta M_1 $ do not contribute to (\ref{diff}) as they lower the degree of divergence of $\int d^4 p \,h^{\nu \rho}(p)$ from quadratic to logarithmic (a shift of the integration variable is allowed in logarithmically divergent integral). A priori, a non vanishing effect can be provided by the terms linear in $\delta M_1$, which give the following contribution to (\ref{hmu})
\bea h^{\nu \rho}_1(p)\equiv \mbox{Tr}\left( \gamma^{\rho}\gamma_5\frac{1}{\displaystyle{\not} p-\displaystyle{\not} \alpha-m-i\epsilon}\delta M_1\frac{1}{\displaystyle{\not} p-\displaystyle{\not} \alpha-m-i\epsilon}\gamma^{\nu}\frac{1}{\displaystyle{\not} p+ \displaystyle{\not}k_1-\displaystyle{\not}\alpha-m-i\epsilon}\right. \nonumber \\  +\left.  \gamma^{\rho}\gamma_5\frac{1}{\displaystyle{\not} p-\displaystyle{\not} \alpha-m-i\epsilon}\gamma^{\nu}\frac{1}{\displaystyle{\not} p+ \displaystyle{\not}k_1-\displaystyle{\not} \alpha-m-i\epsilon}\delta M_1\frac{1}{\displaystyle{\not} p+ \displaystyle{\not}k_1-\displaystyle{\not}\alpha-m-i\epsilon}\right).\eea
However, it is easy to see that the linear divergence cancels in the last expression because $\delta M_1$ contains an even\footnote{Therefore, $\delta M_1$ does not contribute to $\frac{\partial}{\partial x^{\mu}}\langle j^{\mu}_b(x) j^{\nu}_c(y)j^{\rho}_{5d}(z)\rangle_{anom}$ for the same reason why $m$ does not.}  number of Dirac matrices and the trace of an odd number of Dirac matrices vanishes. So $\int d^4 p \,h_1^{\nu \rho}(p)$ is also at most logarithmically divergent and does not contribute to (\ref{diff}). A similar argument shows that the 
$\delta M_1$ dependence in the  second and third terms in the square brackets of (\ref{anomVVA}) disappears. Analogous is also the treatment of the AWIs and therefore we do not explicitly present it here. Moreover, we observe that it is possible to derive in the same way the well-known $m$-independence of the anomaly.

The proof that the antisymmetric part of (\ref{VVA}) satisfies the WIs can also be based on a quite similar reasoning. By writing down an explicit expression for the 2-point functions appearing in the WIs one can easily realize that the condition for the WIs to be satisfied by the antisymmetric part again coincides\footnote{This is indeed true for the choice of the labeling of the momenta discussed below Eq. (\ref{VVA}).} with the possibility to perform a shift of the integration variable in a potentially divergent integral. Since this is allowed in the $\delta M_1=0$ (Lorentz invariant) case, so it is in the  $\delta M_1 \neq 0$ case  because extracting powers of $\delta M_1$ necessarily lowers the degree of divergence.

\section{Anomaly cancellation and regulator independence}\label{regulator-indep}

In the last section we have shown that the anomalous part of the 3-point functions may be taken equal to the Lorentz invariant case. In other words, there are regulators for which the anomalous part is insensitive to the Lorentz violating parameters. It might still be possible that such an anomaly can be canceled by adding a Lorentz violating term $\mathcal{P}[A]$ to $W[A]$,  where $\mathcal{P}[A]$ is a local integral of polynomials in $A_{\mu}$ and its derivatives (counterterms). Here we show that this is not the case and therefore the anomaly cancellation conditions remain at least as strong as in Lorentz invariant theories even if we allow for Lorentz violating regularizations.

From the results of Section \ref{fermionredef} we know that
\bea  &&\frac{\partial}{\partial x^{\mu}}\langle J^{\mu}_{Lb}(x) J^{\nu}_{Lc}(y)J^{\rho}_{Ld}(z)\rangle_{anom}=\nonumber \\ &&-\frac{\pi^2}{(2 \pi)^{12}}\mbox{Tr}\left(t^b_L \{t^c_L,t^d_L\}\right) \int d^4 k_1 d^4 k_2 e^{i\left[\left(k_1+k_2\right)x-k_1 y - k_2 z\right]} \epsilon^{\kappa \nu \lambda \rho} \eta_{\kappa} (k_1+k_2)_{\lambda},\label{anometa}\eea
where $\eta_{\mu}$ is defined in terms of $\alpha_{\mu}$ appearing in (\ref{divergence}) by means of the following relation
\be \eta_{\mu}\equiv \alpha_{\mu} - l^{(-1)\rho}_{\quad \quad\mu}(a_{\rho}-b_{\rho}), \label{etaalpha}\ee
with $l^{(-1)\rho}_{\quad \quad\mu} l^{\mu}_{\,\,\, \tau}= \delta^{\rho}_{\tau}$. To obtain (\ref{anometa}) we used together Eqs. (\ref{LV-LIanom}), (\ref{property}) and (\ref{explicit-anom}) and redefinitions (\ref{shift-redef}) and (\ref{Red2}). Therefore, like in the Lorentz invariant case (see Eq. (\ref{LIdeltaW})), the anomalous part of $\delta W[A]$, which is quadratic in the gauge fields, can be written as follows:
\be \delta W[A]_{\,anom}^{(2)} = \frac{1}{48\pi^2}  \mbox{Tr}\int d^4x  \,\epsilon^{\mu \nu \lambda \rho} \Omega_L \,\partial_{\mu} \left(2 A_{\nu}^{L} \partial_{\lambda}A_{\rho}^L\right)- (L\rightarrow R), \label{2deltaW}\ee
where label $(2)$ denotes the part quadratic in $A_{\mu}$. Regarding the left-handed part\footnote{There are analogous relations for the right-handed part.}, in the Lorentz invariant case this corresponds to choosing $\alpha_{\mu}=(k_{1\mu}-k_{2\mu})/3$ in (\ref{explicit-anom}), whereas in the general Lorentz violating case to choosing 
\be \alpha_{\mu}=l^{(-1)\rho}_{\quad \quad\mu}(a_{\rho}-b_{\rho})+\frac{1}{3}(k_{1\mu}-k_{2\mu}) \ee
in Eq. (\ref{anometa}), where we used (\ref{etaalpha}).

There are two types of counterterms, which after a gauge variation (\ref{Atransf}) could cancel $\delta W[A]_{\,anom}^{(2)}$:
\begin{enumerate}
 \item terms cubic in the gauge fields and with one derivative,
\item terms quadratic in the gauge fields with two derivatives.
\end{enumerate}
Therefore, a counterterm that could cancel the part of the anomaly quadratic in $A_{\mu}$ is
\bea \mathcal{P}[A]=\int d^4x \left(c_{bcd}^{\mu \nu \rho \sigma}\, \partial_{\mu}A^b_{\nu}A^c_{\rho} A^d_{\sigma}+ q_{(1)bc}^{\mu \nu \rho \sigma}\, \partial_{\mu}\partial_{\nu}A^b_{\rho} A^c_{\sigma}+q_{(2)bc}^{\mu \nu \rho \sigma} \,\partial_{\mu}A^b_{\nu}\partial_{\rho} A^c_{\sigma}\right), \label{counterterms}\eea
where $c_{bcd}^{\mu \nu \rho \sigma} , q_{(1)bc}^{\mu \nu \rho \sigma}$ and $q_{(2)bc}^{\mu \nu \rho \sigma} $ are constant, but not assumed to be Lorentz invariant tensors.
Let us start by analyzing the first term in (\ref{counterterms}); after a gauge variation (\ref{Atransf}) this gives
\bea &&\delta \left( c_{bcd}^{\mu \nu \rho \sigma}\,\int d^4x \,\partial_{\mu}A^b_{\nu}A^c_{\rho} A^d_{\sigma}\right)=  c_{bcd}^{\mu \nu \rho \sigma} \int d^4 x\,\left[\Omega^b \partial_{\mu}\partial_{\nu}\left( A_{\rho}^cA_{\sigma}^d\right) \right. \nonumber \\&& \left. - \Omega^c \partial_{\rho}\left(\partial_{\mu} A_{\nu}^b A_{\sigma}^d \right)-\Omega^d\partial_{\sigma}\left(\partial_{\mu} A_{\nu}^b A_{\rho}^c \right) + \mathcal{O}\left(A^3\right) \right],\label{gaugevar}
\eea
where $\mathcal{O}\left(A^3\right)$ represents terms cubic in the gauge fields and we have performed some integrations by parts. If (\ref{gaugevar}) could cancel $\delta W[A]_{\,anom}^{(2)}$, it should be invariant under general coordinate transformations (which are connected to the identity) and odd under parity and time-reversal because (\ref{2deltaW}) has these properties. Therefore, it would be not restrictive to take $c_{bcd}^{\mu \nu \rho \sigma}=T^{bcd} \epsilon^{\mu \nu \rho \sigma}$, where $T^{bcd}$ is arbitrary, in (\ref{gaugevar}) and to start with a Lorentz invariant counterterm in (\ref{counterterms}).  A similar reasoning shows that, if we could cancel $\delta W[A]_{\,anom}^{(2)}$, it would also be not restrictive to take  $q_{(1)bc}^{\mu \nu \rho \sigma}$ and $q_{(2)bc}^{\mu \nu \rho \sigma}$ respectively of the form $T_{(1)}^{bc} \epsilon^{\mu \nu \rho \sigma} $ and $T_{(2)}^{bc} \epsilon^{\mu \nu \rho \sigma} $ and therefore the corresponding terms in $\mathcal{P}[A]$  would reduce to Lorentz invariant counterterms.  

For this reason, in order to cancel the anomalies it is necessary to satisfy the same conditions as in Lorentz invariant theories \cite{Gross:1972pv}
\be \mbox{Tr}\left(t_L^b\{t^c_L,t^d_L\}\right)= \mbox{Tr}\left(t_R^b\{t^c_R,t^d_R\}\right), \ee
because it is required to cancel $\delta W[A]_{\,anom}^{(2)}$. Therefore, it is not possible to allow more general gauge groups and quantum numbers than those permitted in Lorentz invariant theories. Since our discussion is applicable to both gauge and global anomalies, we have also obtained that, whenever a global anomaly is present in a Lorentz invariant theory it is also present in any Lorentz violating extension of the general form considered here, namely that introduced in Ref. \cite{Colladay:1998fq}.
 In these cases, the results of Section \ref{fermionredef} tell us that we can take the anomalous part of the triangle graphs to be equal to the Lorentz invariant case.

\section{Conclusions} \label{Concl}

We have investigated the role of Lorentz symmetry in the perturbative anomalies due to the presence of (not necessarily dynamical) gauge fields. The class of theories that has been considered is quite general even though some assumptions have been made (see Sections \ref{Intro} and \ref{model}). The main result of the paper is that the anomaly cancellation conditions cannot be relaxed by allowing Lorentz violations.

To obtain this, in section \ref{fermionredef} we have first considered the one-loop 3-point functions for the fermion currents derived from an action that contains general Lorentz violating parameters as those in (\ref{Gansatz}) and (\ref{Mdef}). We showed that the anomalous part of the corresponding Ward identities can be taken equal to the Lorentz invariant limit. In case  $m+H^{\mu \nu}\sigma_{\mu \nu}/2=0$ (that we called massless case), this property can be shown by {\it (i)} performing changes of variable in the integrals w.r.t. the momenta  and {\it (ii)} redefining the labeling of the momenta carried by the internal lines. Step {\it (i)} shows the independence of the Lorentz violating parameters in the one-derivative terms in the action (that is $c^{\mu}_{\,\,\,\nu}-\delta^{\mu}_{\nu}$ and $d^{\mu}_{\,\,\,\nu}$) by exploiting the invariance of the anomaly functional under general coordinate transformations, whereas step {\it (ii)} proves the independence of those appearing in the terms without derivatives (that is $a_{\mu}$ and $b_{\mu}$). The divergent integrals  have been regularized  by performing the momentum integration over a finite domain. After step {\it (i)}, the shape of this domain can be chosen in order to  reproduce the symmetric integration procedure, which is often used to compute the anomalies in standard theories.
In a separate part we have considered case  $m+H^{\mu \nu}\sigma_{\mu \nu}/2\neq 0$ (that we referred to as the massive case). The $H^{\mu \nu}$-independence of the anomaly has been proved by extracting powers of $H^{\mu \nu}\sigma_{\mu \nu}/2$ from the internal propagators. Such a procedure lowers the divergence of the integrals enough to render the anomaly insensitive to $H^{\mu \nu}$.  These results may be useful to compute processes, where the anomalies are relevant, in the Lorentz violating extensions considered for example in Ref. \cite{Colladay:1998fq}. 

The method that we have just summarized appears to be applicable to other types of graphs, like the square graphs. The detailed analysis of other Green functions (different from the 3-point and 2-point functions) would be of interest, but has not been provided here as unnecessary to obtain the main result of the paper. Indeed, in Section \ref{regulator-indep}, we have shown that the anomalous part of the Ward identities for the 3-point functions fixes $\delta W[A]_{\,anom}^{(2)}$ (in some regularization) equal to the Lorentz invariant case and no local counterterms (not necessarily Lorentz invariant) in the Lagrangian can cancel this term after a gauge variation. So the anomaly cancellation conditions turn out to be very stable under Lorentz violating perturbations.

Finally, given that here a perturbative method has been used and only gauge fields have been coupled to fermions, an open issue is the investigation of a similar problem, but for the non-perturbative anomalies and for theories involving gravitational interactions.

\vspace{1cm}

{\bf Acknowledgments.} The author gratefully acknowledges valuable discussions with Mikhail Shaposhnikov. This work is supported  by the Tomalla Foundation.


\begin{thebibliography}{99}

\bibitem{Bell:1969ts}
 J.~S.~Bell and R.~Jackiw,
  ``A PCAC puzzle: pi0 $\to$ gamma gamma in the sigma model,''
  Nuovo Cim.\  A {\bf 60} (1969) 47. 
  S.~L.~Adler,
  ``Axial vector vertex in spinor electrodynamics,''
  Phys.\ Rev.\  {\bf 177} (1969) 2426.


\bibitem{Wald:1980nm}
  {\it A very partial list is} R.~M.~Wald,
  ``Quantum Gravity And Time Reversibility,''
  Phys.\ Rev.\  D {\bf 21} (1980) 2742. 
  V.~A.~Kostelecky and S.~Samuel,
  ``Spontaneous Breaking of Lorentz Symmetry in String Theory,''
  Phys.\ Rev.\  D {\bf 39} (1989) 683.
  V.~A.~Kostelecky and S.~Samuel,
  ``Phenomenological Gravitational Constraints on Strings and Higher
  Dimensional Theories,''
  Phys.\ Rev.\ Lett.\  {\bf 63} (1989) 224.
  V.~A.~Kostelecky and S.~Samuel,
  ``Gravitational Phenomenology In Higher Dimensional Theories And Strings,''
  Phys.\ Rev.\  D {\bf 40} (1989) 1886.
  V.~A.~Kostelecky and S.~Samuel,
  ``Photon and Graviton Masses in String Theories,''
  Phys.\ Rev.\ Lett.\  {\bf 66} (1991) 1811.
  V.~A.~Kostelecky and R.~Potting,
  ``CPT and strings,''
  Nucl.\ Phys.\  B {\bf 359} (1991) 545.
  T.~Jacobson, S.~Liberati and D.~Mattingly,
  ``Quantum gravity phenomenology and Lorentz violation,''
  Springer Proc.\ Phys.\  {\bf 98} (2005) 83
  [arXiv:gr-qc/0404067].

\bibitem{Bertolami:1996cq}
   {\it Some examples are the following:} O.~Bertolami, D.~Colladay, V.~A.~Kostelecky and R.~Potting,
  ``CPT violation and baryogenesis,''
  Phys.\ Lett.\  B {\bf 395} (1997) 178
  [arXiv:hep-ph/9612437].
  V.~A.~Kostelecky and M.~Mewes,
  ``Lorentz and CPT violation in neutrinos,''
  Phys.\ Rev.\  D {\bf 69} (2004) 016005
  [arXiv:hep-ph/0309025].
  M.~Libanov, V.~Rubakov, E.~Papantonopoulos, M.~Sami and S.~Tsujikawa,
  ``UV stable, Lorentz-violating dark energy with transient phantom era,''
  JCAP {\bf 0708} (2007) 010
  [arXiv:0704.1848 [hep-th]].

\bibitem{Kostelecky:2000mm}
  V.~A.~Kostelecky and R.~Lehnert,
  ``Stability, causality, and Lorentz and CPT violation,''
  Phys.\ Rev.\  D {\bf 63} (2001) 065008
  [arXiv:hep-th/0012060].

\bibitem{Kostelecky:2001jc}
  V.~A.~Kostelecky, C.~D.~Lane and A.~G.~M.~Pickering,
  ``One-loop renormalization of Lorentz-violating electrodynamics,''
  Phys.\ Rev.\  D {\bf 65} (2002) 056006
  [arXiv:hep-th/0111123].

\bibitem{Colladay:2006rk}
  G.~de Berredo-Peixoto and I.~L.~Shapiro,
  ``On the renormalization of CPT/Lorentz violating QED in curved space,''
  Phys.\ Lett.\  B {\bf 642} (2006) 153
  [arXiv:hep-th/0607109].
  D.~Colladay and P.~McDonald,
  ``One-loop renormalization of pure Yang-Mills with Lorentz violation,''
  Phys.\ Rev.\  D {\bf 75} (2007) 105002
  [arXiv:hep-ph/0609084].
  D.~Anselmi and M.~Halat,
  ``Renormalization of Lorentz violating theories,''
  Phys.\ Rev.\  D {\bf 76} (2007) 125011
  [arXiv:0707.2480 [hep-th]].
  D.~Colladay and P.~McDonald,
  ``One-Loop Renormalization of QCD with Lorentz Violation,''
  Phys.\ Rev.\  D {\bf 77} (2008) 085006
  [arXiv:0712.2055 [hep-ph]].



\bibitem{Coleman:1998ti}
  S.~R.~Coleman and S.~L.~Glashow,
  ``High-energy tests of Lorentz invariance,''
  Phys.\ Rev.\  D {\bf 59} (1999) 116008
  [arXiv:hep-ph/9812418].


\bibitem{Colladay:1996iz}
  D.~Colladay and V.~A.~Kostelecky,
  ``CPT violation and the standard model,''
  Phys.\ Rev.\  D {\bf 55} (1997) 6760
  [arXiv:hep-ph/9703464].

\bibitem{Colladay:1998fq}
  D.~Colladay and V.~A.~Kostelecky,
  ``Lorentz-violating extension of the standard model,''
  Phys.\ Rev.\  D {\bf 58} (1998) 116002
  [arXiv:hep-ph/9809521].

\bibitem{Arias:2007xt}
  P.~Arias, H.~Falomir, J.~Gamboa, F.~Mendez and F.~A.~Schaposnik,
  ``Chiral Anomaly Beyond Lorentz Invariance,''
  Phys.\ Rev.\  D {\bf 76} (2007) 025019
  [arXiv:0705.3263 [hep-th]].


\bibitem{Bilal:2008qx}
  {\it See for example} A.~Bilal,
  ``Lectures on Anomalies,''
  arXiv:0802.0634 [hep-th].


\bibitem{Zumino:1983rz}
  B.~Zumino, Y.~S.~Wu and A.~Zee,
  ``Chiral Anomalies, Higher Dimensions, And Differential Geometry,''
  Nucl.\ Phys.\  B {\bf 239} (1984) 477.



\bibitem{Wess:1971yu}
  J.~Wess and B.~Zumino,
  ``Consequences of anomalous Ward identities,''
  Phys.\ Lett.\  B {\bf 37} (1971) 95.

\bibitem{Treiman:1986ep}
  S.~B.~Treiman, E.~Witten, R.~Jackiw and B.~Zumino,
  ``Current algebra and anomalies,''
{\it  Singapore, Singapore: World Scientific (1985) 537p}.

\bibitem{Weinberg:1996kr}
  S.~Weinberg,
  ``The quantum theory of fields. Vol. 2: Modern applications,''
{\it  Cambridge, UK: Univ. Pr. (1996) 489p}


\bibitem{Bardeen:1969md}
  W.~A.~Bardeen,
  ``Anomalous Ward identities in spinor field theories,''
  Phys.\ Rev.\  {\bf 184} (1969) 1848.

\bibitem{Gross:1972pv}
  D.~J.~Gross and R.~Jackiw,
  ``Effect of anomalies on quasirenormalizable theories,''
  Phys.\ Rev.\  D {\bf 6} (1972) 477.


\end{thebibliography}
\end{document}